# Phonon drag as a mechanism of delayed terahertz response of metals

*Ivan Oladyshkin\**

*Department of Nonlinear Electrodynamics, A.V. Gaponov-Grekhov Institute of Applied Physics of the Russian Academy of Sciences, 603950, 46 Ul'yanov Street, Nizhny Novgorod, Russia*

\*[oladyshkin@ipfran.ru](mailto:oladyshkin@ipfran.ru)

We show that electron drag by nonequilibrium phonons describes the actual waveform and spectrum of terahertz pulses generated during femtosecond laser irradiation of metals. In contrast to previous models, there is a picosecond delay in the drag force development due to the relatively slow lattice heating and finite phonon lifetime. We also predict that, at high pump fluences, a macroscopic deformation wave enhances nonlinearly the drag force and terahertz response. Our results establish the terahertz pulse waveform as a direct probe of ultrafast lattice dynamics in metals.

Probing ultrafast lattice dynamics in solids on picosecond timescales remains an experimental challenge. In this Letter, we show that laser-induced terahertz (THz) emission from metals offers a direct window into these processes. We demonstrate that the observed waveforms of THz pulses are shaped by the phonon drag force – the delayed transfer of phonon momentum to electrons – which naturally imprints the thermal evolution of the crystal lattice onto the emitted radiation. This makes the THz signal a quantitative probe of picosecond dynamics of heat flux or macroscopic deformation wave at high pump fluences.

Despite previous attempts to interpret the THz response of metals [1–5], the most essential features of this effect remain unexplained from a microscopic point of view. The key problem is the universally delayed character of the THz signal: a femtosecond laser pump generates a much longer low-frequency pulse with the duration of 1–2 ps and the spectral range of 0.3–2 THz [6-13].

In experimental papers [6–8], where optical-to-THz conversion on metal surfaces was first observed, only a phenomenological description of this effect was provided in terms of nonlinear susceptibility $\chi_2$. The first microscopic models [1–3], based on the ponderomotive effect significantly underestimated the actual THz amplitude and predicted direct laser pulse rectification, which contradicted experimental observations. Subsequent microscopic models invoked the gradient of hot-electron pressure and the photon drag force as the THz radiation sources [3–5]. These two sources are comparable in magnitude [5] and significantly exceed the ponderomotive force. Despite a slight delay in electron temperature dynamics, the predicted THz signal, however, was still close to the laser pulse envelope.

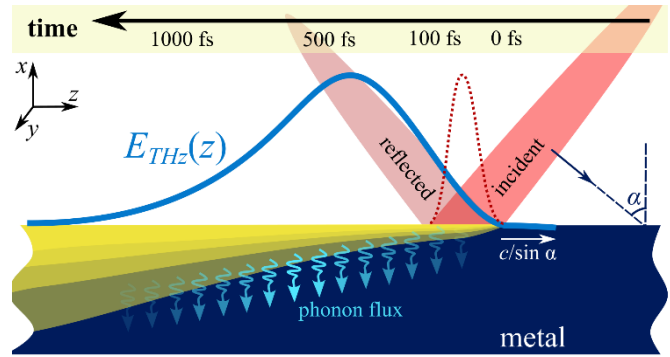

**Fig. 1.** Schematic of phonon-driven THz generation. Yellow gradient is the lattice temperature distribution; wavy blue arrows depict the phonon flux. The thick blue curve shows the THz field $E_z(z)$, produced by phonon drag at the metal surface, while the dotted red curve is the laser pulse envelope. The timeline at the top marks the time after the pump arrival.

Nevertheless, a noticeable outcome of these studies is the introduction of a Cherenkov interaction geometry for the case of oblique pump incidence, when THz field is generated by a light spot moving along the metal surface at a superluminal phase velocity (see Fig. 1). This framework successfully explained the THz radiation direction coinciding with the pump reflection. Also, in these papers several electrodynamical problems of low-frequency radiation from various nonlinear sources were solved, which will be used in the present Letter in a generalized form.

The models discussed above assumed that free electrons experience a classical "viscous friction" force from the crystal lattice, and that momentum transferred to the lattice is irreversibly lost. This assumption is, however, generally invalid: upon collision, electron momentum is transferred to the phonon subsystem and may, with a certain probability, return to the electronic subsystem via subsequent collisions. Phenomena of this kind are termed *phonon wind* or *phonon drag* when they involve the exchange of directed momentum, and *thermal drag* when the force originates from a temperature gradient in the phonon gas [14, 15]. The eventual dissipation of momentum of the electron-phonon system occurs only on picosecond timescales, predominantly through Umklapp phonon scattering, whereby momentum is transferred to the crystal lattice as a whole. In stationary regimes, *e-ph* momentum exchange merely renormalizes the effective electron-phonon collision frequency. But on picosecond timescales, when the two subsystems are far from equilibrium, momentum exchange can induce a pronounced temporal dispersion of the electronic response.

Consider the phonon drag effect in the case when the crystal lattice of metal is nonuniformly heated due to the laser pulse absorption. There is a directed phonon flux that transports heat away from the near-surface layer, and the momentum of phonons can be transferred to electrons during collisions. To make preliminary estimates, we assume the diffusive regime of heat transport, when the phonon heat flux, $\mathbf{\Pi}_{ph}$, is governed by Fourier's law:

$$\mathbf{\Pi}_{ph} = -\varkappa_l \nabla T_l, \quad (1)$$

where $\varkappa_l$ and $T_l$ are the crystal lattice thermal conductivity and temperature, respectively. Microscopically, the energy flux $\mathbf{\Pi}_{ph}$ is created by propagating acoustic phonons with the dispersion law $\omega = q c_s$, where $q$ and $\omega$ are the phonon wavenumber and frequency, respectively, and $c_s$ is the sound velocity. Here we neglect the differences in sound velocity among the various phonon modes and deviations from linear dispersion, as they are presently inessential. The energy flux of acoustic phonons is linearly related to their macroscopic momentum: $\mathbf{\Pi}_{ph} = c_s^2 \mathbf{P}_{ph}$, where $\mathbf{P}_{ph}$ is the momentum of phonons per unit volume.

Introducing the phonon-electron scattering rate $\nu_{ph-e}$, we can express the phonon drag force $\mathbf{F}_{ph}$ acting on the electronic subsystem in the unit volume:

$$\mathbf{F}_{ph} = \nu_{ph-e} \mathbf{P}_{ph} = -\frac{\varkappa_l \nu_{ph-e}}{c_s^2} \nabla T_l. \quad (2)$$

According to Eq. (2), the temporal evolution of the drag force depends directly on the dynamics of the crystal lattice temperature, and so, it does not follow the laser pulse envelope. In the relaxation-time approximation, the lattice thermal conductivity takes the standard form

$$\varkappa_l = \frac{c_s^2}{3 \nu_{ph}} C_l, \quad (3)$$

where $\nu_{ph}$ is the total scattering frequency of phonons and $C_l$ is the heat capacity of the crystal lattice per unit volume. Therefore, we obtain:

$$\mathbf{F}_{ph} = -\frac{1}{3} \frac{\nu_{ph-e}}{\nu_{ph}} \nabla U_l, \quad (4)$$

where $U_l = C_l T_l$ is the thermal energy of the crystal lattice. The ratio $\nu_{ph-e}/\nu_{ph}$ quantifies the fraction of the phonon momentum that is returned to the electrons and typically lies in the range of 0.1–0.9, depending on the strength of electron-phonon coupling in a specific metal [16, 17].

Eq. (4) allows us to compare directly the phonon drag force with the electronic pressure gradient $\mathbf{F}_e$:

$$\mathbf{F}_e = -\frac{2}{3}\nabla U_e, \tag{5}$$

where $U_e$ is the average kinetic energy of electrons per unit volume. One can see that $\mathbf{F}_{ph}$ and $\mathbf{F}_e$ have similar gradient form, and their amplitudes depend on the distribution of thermal energy between the electrons and the crystal lattice.

At the initial stage of interaction, the laser pulse energy is deposited in the electron subsystem, so initially $\mathbf{F}_e$ will exceed $\mathbf{F}_{ph}$, especially if the heat exchange between electrons and crystal lattice is slow (like in Au). During further thermalization almost all absorbed energy will be transferred to the crystal lattice since its heat capacity is much larger than the heat capacity of electron gas. At this stage, the phonon drag force (4) will exceed electronic pressure (5) by a factor of several to several tens, depending on $\nu_{ph-e}/\nu_{ph}$. In metals with strong electron-phonon coupling and fast thermalization (like Fe, Ni, Co), the phonon drag force is expected to exceed the electronic pressure force even during the laser pulse action.

Introduction of the phonon drag force $\mathbf{F}_{ph}$, which is substantially delayed relative to the laser pump, allow us, in principle, to interpret the actual waveforms of THz pulses. However, aiming to describe correctly the phonon flux dynamics at picosecond timescales, we need to improve the model of phonon transport. The classical heat conductivity model, used for the above estimates, assumes that the energy flux (1) appears instantaneously with the lattice temperature gradient. But it is valid only at timescales larger than the time of phonon mean free path $\tau_{ph} = \nu_{ph}^{-1}$, which is typically equal to several picoseconds [16–18]. At shorter timescales, the finite phonon lifetime $\tau_{ph}$ cannot be neglected.

To account for the finite phonon lifetime, we employ the Boltzmann kinetic equation for the phonon distribution function. Following the standard moment method in the relaxation-time approximation [14] and retaining the explicit time derivatives essential for the picosecond dynamics, we obtain the following equations for the total phonon momentum $\mathbf{P}_{ph}(\mathbf{r}, t)$ and total phonon energy $U_l(\mathbf{r}, t)$ per unit volume (see the derivation in **Supplementary Materials, Section A**):

$$\frac{\partial \mathbf{P}_{ph}}{\partial t} + \frac{1}{3}\nabla U_l = -\nu_U \mathbf{P}_{ph} - \nu_{ph-e}\mathbf{P}_{ph} + \nu_{e-ph}\mathbf{P}_e, \tag{6}$$

$$\frac{\partial U_l}{\partial t} + c_s^2 \, div \, \mathbf{P}_{ph} = S_e, \tag{7}$$

where $S_e$ describes the crystal lattice heating by electrons, $\mathbf{P}_e$ is the momentum of electrons per unit volume, $\nu_U$ is relaxation rate due to U-processes, $\nu_{e-ph}$ and $\nu_{ph-e}$ are the frequencies of *e-ph* and *ph-e* collisions, respectively. The frequencies $\nu_{e-ph}$ and $\nu_{ph-e}$ obey the balance relation $\nu_{e-ph}n_e = \nu_{ph-e}n_{ph}$, where $n_{ph}$ and $n_e$ are the concentrations of phonons and free electrons, respectively (at room temperature $n_{ph}$ exceeds $n_e$ by 1–2 orders of magnitude, so the phonon lifetime is much larger than the electron free path time). In contrast to quasi-stationary treatments of phonon drag, Eqns. (6) and (7) capture the transient buildup of the phonon flux, which is responsible for the delayed THz response.

The relaxation rates governing the phonon dynamics in our model were analyzed in detail in Refs. [16–18]. According to [18], the energy transfer from hot electrons to the lattice proceeds predominantly into longitudinal acoustic (LA) phonons, while transverse acoustic (TA) phonons are populated indirectly via anharmonic phonon–phonon decay on a timescale of ~0.2–0.5 ps. The full thermalization of the phonon subsystem, and the corresponding dissipation of the directed phonon momentum, occurs within 1–3 ps due to Umklapp scattering. This naturally introduces the picosecond delay required to explain the observed THz pulse durations.

Equations (6)–(7) describe the dynamics of the phonon flux coupled with electrons, accounting explicitly for the finite phonon lifetime. The phonon flux does not arise instantaneously with the crystal lattice heating, while its growth rate is proportional to the gradient of the lattice thermal energy $\nabla U_l$. The stationary flux is established during the time $t_{ph} \approx (\nu_U + \nu_{ph-e})^{-1}$, i.e., in several picoseconds. At timescales $t \gg t_{ph}$ this system describes the classical regime of diffusive thermal conductivity.

Since the phonon drag effect is an integral characteristic of the phonon distribution function, here we use the two-temperature model for $S_e$ [19, 20], assuming that the heat flux arises on the background of locally-thermalized distribution of phonons $f_0$ with the temperature $T_l$. Although the electron–phonon interaction is mode-dependent [16–18], the two-temperature model accurately captures the integrated energy exchange [16]. At low pump fluences the relaxation rates can be considered constant, but, in the general case, their dependence on temperatures of electrons and lattice must be taken into account.

In two-temperature model, the energy transfer from the electrons to the crystal lattice is described by the following equation:

$$S_e = \mathrm{G} C_e (T_e - T_l), \qquad (8)$$

where $C_e \cong n_e \frac{\pi^2}{2} \frac{k^2 T_e}{E_F}$ is the heat capacity of degenerate electron gas per unit volume ($k$ is the Boltzmann constant, $E_F$ is the Fermi energy), $\mathrm{G}$ is the temperature relaxation rate. For the evolution of electron temperature, we can write the diffusion-type equation:

$$\frac{\partial T_e}{\partial t} = D_e \frac{\partial^2 T_e}{\partial x^2} - \mathrm{G}(T_e - T_l) + S_{opt}, \qquad (9)$$

where $D_e$ is the electron thermal diffusivity and $S_{opt}$ is the optical heat source (localized in the skin-layer). Here $D_e$ can be expressed as $D_e = v_F^2 \tau_e / 3$, where $v_F$ is the Fermi velocity and $\tau_e$ is the time of electron free path.

Finally, to close the system, we add the equation of electron motion:

$$\frac{\partial \mathbf{P}_e}{\partial t} = -e n_e \mathbf{E} + \nu_{ph-e} \mathbf{P}_{ph} - \nu_{e-ph} \mathbf{P}_e, \qquad (10)$$

where $\mathbf{E}$ is the self-consistent electric field inside the metal. The terms $\nu_{ph-e} \mathbf{P}_{ph}$ and $\nu_{e-ph} \mathbf{P}_e$, which appear symmetrically in Eqns. (6) and (10), describe the drag force between the electrons and phonons.

For brevity, we do not include the gradient of electron pressure in Eq. (10), since the problem of radiation from this source has already been solved [4, 5] (see also **Supplementary Materials, Section B**). We also omit explicit treatment of electron collisions with impurities and defects, since electron–phonon scattering dominates at room temperature and above. Finally, we neglect *e-e* Umklapp scattering, which contributes significantly to momentum dissipation only when the electron temperature is comparable with the Fermi energy.

To relate the resulting model to the THz emission, one must solve the problem of radiation from electrons driven by the phonon drag force. Similar problems for various sources have already been thoroughly analyzed in [1–6]; the relevant results are summarized in **Supplementary Materials, Section B**. Briefly, because the plasma frequency in metals is extremely high ($\sim 10^{16}\,\mathrm{s}^{-1}$), any potential force acting on the electron subsystem is screened almost instantaneously by charge-separation field $\mathbf{E}$ ($rot\,\mathbf{E} \cong 0$ since the field is quasistatic). Consequently, the electric current is negligible, so that $\mathbf{P}_e \cong 0$. Radiation from this system arises solely due to the presence of the boundary and the slow variations of the quasistatic fields.

The phonon drag force $\mathbf{F}_{ph} = \nu_{ph-e} \mathbf{P}_{ph}$, driven by inhomogeneous heating of the crystal lattice, is also a potential force and is therefore compensated by a quasistatic charge-separation field $\mathbf{E}$ inside the metal:

$$e n_e \mathbf{E} = \nu_{ph-e} \mathbf{P}_{ph}. \qquad (11)$$

Using this relation, one can rewrite Eq. (8) as the equation on $\mathbf{E}$:

$$\frac{\partial \mathbf{E}}{\partial t} + \nu_{ph}\mathbf{E} = -\frac{\nu_{ph-e}}{3en_e}\nabla U_l, \tag{12}$$

where $\nu_{ph} = \nu_{ph-e} + \nu_U$ is the total collision rate of phonons. Eq. (12) allows us to express analytically the electric field **E**:

$$\mathbf{E}(\mathbf{r},t) = -\frac{1}{3}\frac{\nu_{ph-e}C_l}{en_e}\int_{-\infty}^{t}\nabla T_l(\mathbf{r},t')e^{\nu_{ph}(t'-t)}dt'. \tag{13}$$

In its spatial structure, $\mathbf{E}(\mathbf{r},t)$ resembles that of an electrical double layer with a time-varying surface charge $\sigma(z,t)$. The normal component of the field is localized inside the metal: $E_x = 0$ at the boundary since $\frac{\partial T_l}{\partial x} = 0$ (no heat flux across the surface). The tangential component $E_z$, however, is nonzero and continuous across the boundary, so it can be used to find low-frequency radiation.

The most transparent analytical solution can be obtained for the Cherenkov geometry, which is typical for real experiments. Consider a laser pulse obliquely incident on the metal at an angle $\alpha$, so that the THz signal is generated within a narrow interaction region that moves along the surface with a superluminal phase velocity $c/\sin\alpha$ (see Fig. 1). If the laser spot diameter is large compared to the THz wavelength, all distributions become stationary in the reference frame comoving with the interaction region. Mathematically, this translates into the relation $\frac{\partial}{\partial t} = -\frac{c}{\sin\alpha}\frac{\partial}{\partial z}$.

Under these conditions, Eq. (13) yields the following expression:

$$E_z(t) = \frac{\sin\alpha}{c}\frac{\nu_{ph-e}C_l}{3en_e}\int_{-\infty}^{t}\frac{\partial T_l}{\partial t'}e^{\nu_{ph}(t'-t)}dt', \tag{14}$$

which gives the explicit relation between the tangential low-frequency electric field and the "history" of the crystal lattice temperature $T_l(\mathbf{r},t')$ at a given point, including the metal boundary $x = 0$.

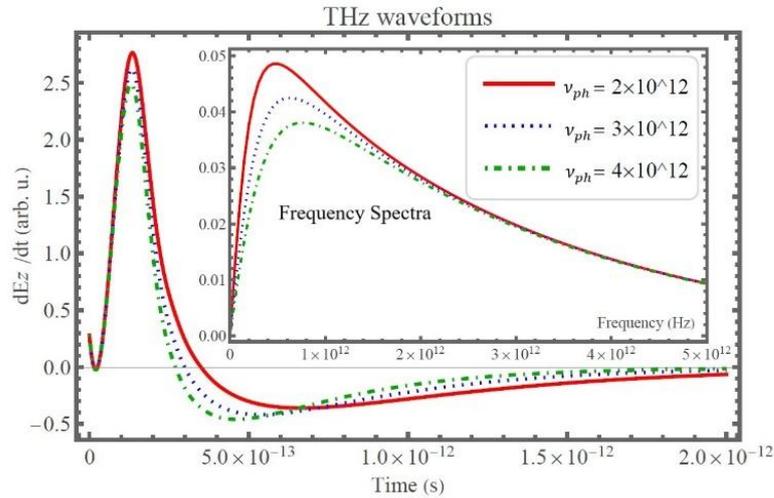

**Fig. 2.** Waveforms of THz pulses in far-field zone $dE_z(t)/dt$, calculated for the parameters of Au and for different $\nu_{ph}$. Inset shows the signals' spectra. The pump FWHM is 100 fs.

There are two important limits of Eq. (14). In the phonon relaxation time $\nu_{ph}^{-1}$ is much larger than the characteristic time of lattice temperature variation, we obtain $E_z \propto (T_l(t) - T_0)e^{-\nu_{ph}t}$, where $T_0$ is the initial lattice temperature. In the opposite case $E_z \propto \frac{1}{\nu_{ph}}\frac{\partial T_l(t)}{\partial t}$, which corresponds to the diffusion limit (Eqns. (2) and (3)). Thus, the phonon drag force produces THz pulse with the waveform determined by the dynamics of the crystal lattice temperature $T_l(t)$ and by the relaxation of phonons.

Under chosen geometry of oblique incidence, the magnetic field $H_y$ on the metal surface is given by $H_y = -E_z/\cos\alpha$, which allow us to find the Poynting vector and estimate the total THz pulse energy for given size of the laser spot.

In Fig. 2, the waveforms and spectra of THz pulses, obtained from Eq. (14) and two-temperature model (10)–(13), are shown. Depicted graphs of $dE_z/dt$ describe the THz signal waveforms in far-field zone, and the chosen range of $\nu_{ph} = (2 \div 4) \cdot 10^{12} \, s^{-1}$ corresponds to the lifetime of TA-phonons calculated in [18]. In contrast to the models based on purely electron response, the phonon drag mechanism predicts THz pulses of picosecond duration, with the spectral maximum in the range of 0.3–1 THz, depending on the phonon relaxation rate $\nu_{ph}$ and the thermalization rate G. These results are in good agreement with those obtained for the bulk Au [7], Au and Ag films [8, 10], Au gratings [9] and Fe [6].

We also note that the enhanced THz emission observed from grating-structured surfaces [9, 12, 13] is naturally explained by the resonant absorption increase [21], while the underlying phonon-drag mechanism remains the same. Conversely, the absence of THz generation from thin gold films (≲100 nm) reported in [8] follows from the absence of a directed phonon flux in the uniformly heated film.

The analysis above assumes a phonon distribution close to thermal equilibrium. At pump fluences of about 0.1 J/cm² and higher, however, the situation changes dramatically: the overheated region emits a coherent deformation wave, which can evolve into a shock wave. As shown in molecular dynamics simulations and experiments (see, for example, [22, 23]), at absorbed fluences of 0.5–1 J/cm², such a deformation wave accumulates tens of percent of the absorbed energy.

In terms of phonons, this can be viewed as the stimulated excitation of a selected phonon mode against the background of spontaneous excitation of the entire phonon spectrum, i.e., lattice heating. Microscopically, the stimulated emission originates from the fact that the probability of emitting of a phonon with a given wavevector **q** is proportional to the population of that mode [16, 17].

Let us estimate the drag force for this case, using macroscopic mechanical treatment. Consider the near-surface layer of the crystal lattice receiving the thermal energy $\Delta U_l = \Phi_{abs}/l_d$, where $\Phi_{abs}$ is the absorbed fluence and $l_d$ is the heat diffusion depth. The pressure increase $\Delta P$ is given by $\Delta P = \Gamma \Delta U_l$, where $\Gamma$ is the Grüneisen parameter, $\Gamma = (1 \div 3)$ for metals. The pressure relaxation takes time $\tau_s = l_d/c_s$, and during this time the layer of the density $\rho$ obtains the velocity $v_P \cong \Delta P / \rho c_s$. Its kinetic energy per unit volume is given by

$$E_K = \frac{1}{2\rho}\left(\frac{\Delta P}{c_s}\right)^2. \tag{15}$$

The fraction of absorbed energy converted into kinetic energy is therefore

$$\eta = \frac{E_K}{\Delta U_l} = \frac{\Gamma^2}{2}\frac{\Phi_{abs}}{\rho c_s^2 l_d}. \tag{16}$$

For the parameters of Au, Al and Cu and for $\Phi_{abs} \approx 0.1 \, J/cm^2$ we find $\eta \approx 10 \div 30\%$, which is in good agreement with [22, 23].

Transformation of the pump energy directly into the deformation wave enhances nonlinearly the phonon drag effect. The momentum carried by the deformation wave, $P_K = E_K/c_s$, is transferred to electrons with the frequency $\nu_{ph-e}$, creating the drag force $F_K = \nu_{ph-e} P_K$. Compare $F_K$ with the drag force produced by phonon diffusion (4):

$$\beta = \frac{F_K}{F_{ph}} = \frac{3\Gamma^2}{2}\frac{\nu_{ph}\Phi_{abs}}{\rho c_s^3}. \tag{17}$$

Estimates for Au, Al and Cu give similar results and show that the influence of the deformation wave begins to dominate ($\beta > 1$) at the absorbed fluences of $10 - 30 \, mJ/cm^2$.

Finally, the condition of the quasistatic electron equilibrium (11) and $rot\ \mathbf{E} \cong 0$ allows us to estimate the tangential electric field $E_z$ generated at the metal surface:

$$E_z \cong \frac{\Gamma^2}{2} \frac{1}{en_e} \frac{\nu_{ph-e} \Phi_{abs}^2}{\rho c_s^3 l_d} \frac{2\pi}{\lambda_{THz}}, \tag{18}$$

where $\lambda_{THz} \approx 0.3\ mm$ is the THz wavelength. Note that in this regime the THz pulse energy scales as the fourth power of the absorbed fluence. Owing to the nonlinearity of laser absorption, the dependence on the incident fluence is expected to be even steeper.

Apparently, this regime of THz generation from Cu was observed in [11, 24], where the directly measured THz pulse energy reaches 0.1–0.2 nJ at pump fluences up to 1 J/cm². For the parameters of Cu and $\Phi_{abs} = 0.2\ J/cm^2$, from Eq. (18) we obtain $E_z \sim 2\ kV/cm$. Accounting for the geometry of the laser beam focusing, described in [24], we find the THz pulse energy ~0.3 nJ, which is close to the measured one.

In conclusion, we have proposed a mechanism for optical-to-THz conversion on metal surfaces, based on the phonon drag force. The phonon flux develops on picosecond timescales with the crystal lattice heating, which naturally explains the significant delay in the THz response and the actual spectrum of THz radiation. At high pump fluences, a macroscopic deformation wave starts to dominate over the diffusive phonon flux, which nonlinearly enhances the drag effect and modifies the THz generation regime. We emphasize that the proposed model relies on the constants of *e–ph* and *ph–ph* interactions, and therefore admits quantitative comparison with experiment, limited only by the accuracy with which these parameters are known form measurements or first-principle calculations.

**Acknowledgements**

The author is grateful to A. V. Korzhimanov, E. D. Gospodchikov, V. A. Mironov and M. D. Tokman for fruitful discussions.

**Funding**

The work was supported by the Russian Science Foundation, grant #25-22-20019.

**Conflict of interests**

The author declares no conflict of interest.

# Supplementary Materials

**A. Derivation of the momentum equations for the Boltzmann equation**

We introduce the phonon distribution function over momenta $f_{ph}(\mathbf{q}, \mathbf{r}, t)$. In this section we assume that only a small fraction of phonons participate in the formation of the directed energy flux, so that the distribution function can be treated perturbatively:

$$f_{ph}(\mathbf{q}, \mathbf{r}, t) = f_0(q, \mathbf{r}, t) + \delta f(\mathbf{q}, \mathbf{r}, t), \tag{S1}$$

where $f_0$ is the isotropic part of the distribution, describing the total energy of the phonon subsystem (or its effective temperature) and depending only on $q = |\mathbf{q}|$, while $\delta f(\mathbf{q})$ is a small anisotropic perturbation that describes the phonon flux.

The time evolution of the full distribution function is governed by the Boltzmann kinetic equation:

$$\frac{\partial f_{ph}}{\partial t} + \frac{\partial f_{ph}}{\partial \mathbf{r}} \mathbf{c}_s = s_e(q, \mathbf{r}, t) + St_{th}\{f_0\} + St_U\{\delta f\} + St_{e-ph}\{f_e, f_{ph}\}, \tag{S2}$$

where $s_e(q, \mathbf{r}, t)$ is the isotropic (in $\mathbf{q}$) source of phonon generation by hot electrons, $\mathbf{c}_s$ is the phonon velocity vector, and $St_i$ are the collision integrals. Specifically, $St_{th}\{f_0\}$ governs thermalization within the phonon subsystem via normal phonon–phonon collisions that conserve total momentum and energy; $St_U\{\delta f\}$ represents momentum loss through Umklapp ph–ph scattering (U-processes); and $St_{e-ph}\{f_e, f_{ph}\}$ describes momentum exchange with the electron subsystem characterized by the distribution function $f_e$ (its contribution is discussed below). The explicit form of the first collision integral is not required, while the second one is treated in the relaxation-time approximation as $t_m\{\delta f\} = -\nu_U \delta f$, where $\nu_U^{-1}$ is the characteristic momentum relaxation time due to U-processes.

Of course, the relevant relaxation rates depend on the phonon energy and on the electron temperature; their behavior is known primarily from theoretical calculations (see, e.g., Refs. [16–18], *references are numbered as in the main text*). Typically, high-energy phonons with wave numbers approaching the Brillouin zone boundary exhibit scattering rates an order of magnitude larger than those of phonons with energies around 0.02–0.03 eV, corresponding to room temperature. In what follows, we work with the moments of $f_{ph}$ (the average momentum and energy of phonons per unit volume). Since the lattice is strongly heated in the regime of interest, the relevant collision frequencies are taken to be those characteristic of high-energy phonons, which dominate the momentum exchange.

Under these assumptions, the evolution of the anisotropic perturbation $\delta f$ is governed by:

$$\frac{\partial \delta f}{\partial t} + \frac{\partial f_0}{\partial \mathbf{r}} \mathbf{c}_s = -\nu_U \delta f + St_{e-ph}. \tag{S3}$$

Here we have used the fact that the collision integral $St_{th}\{f_0\}$ and the source $s_e(q, \mathbf{r}, t)$ are isotropic in momentum and thus do not contribute to $\delta f$. The total phonon momentum per unit volume, $\mathbf{P}_{ph}$, and the total phonon energy per unit volume, $U_l$ (the thermal energy of the crystal lattice), are expressed in terms of $\delta f$ and $f_0$ as

$$\mathbf{P}_{ph} = \int \delta f \, \hbar \mathbf{q} \, d^3 \mathbf{q}, \tag{S4}$$

$$U_l = \int f_0 \, \hbar q c_s \, d^3 \mathbf{q}. \tag{S5}$$

To find the equation of $\mathbf{P}_{ph}$ change, multiply the kinetic equation (S3) by $\hbar \mathbf{q}$ and integrate it over the whole $\mathbf{q}$-space:

$$\frac{\partial \mathbf{P}_{ph}}{\partial t} + \frac{1}{3} \nabla U_l = -\nu_U \mathbf{P}_{ph} + \int St_{e-ph} \hbar \mathbf{q} \, d^3 \mathbf{q}. \tag{S6}$$

Here we have taken into account that the distribution function $f_0$ is isotropic, so the integrals of the form $\int f_0 q_i^2\, d^3\mathbf{q}$ are equal to each other for any projection $q_i = \{q_x, q_y, q_z\}$.

The last term in Eq. (S6) describes momentum exchange between the phonon and electron subsystems. Evidently, in the absence of directed momentum in either subsystem, this term must vanish, irrespective of whether the subsystems are thermalized. A nonzero value of the integral can only arise from anisotropies in the electron and phonon distribution functions, which, for small perturbations, are linearly related to the momentum of the respective subsystem. In particular, if the phonon subsystem carries momentum $\mathbf{P}_{ph}$, it transfers momentum to the electrons at a rate $\nu_{ph-e}\mathbf{P}_{ph}$; similarly, if the electron subsystem carries momentum $\mathbf{P}_e$, it is transferred to the phonons at a rate $\nu_{e-ph}\mathbf{P}_e$, where $\nu_{e-ph}$ is the electron–phonon collision frequency. We thus obtain:

$$\frac{\partial \mathbf{P}_{ph}}{\partial t} + \frac{1}{3}\mathbf{\nabla} U_l = -\nu_U \mathbf{P}_{ph} - \nu_{ph-e}\mathbf{P}_{ph} + \nu_{e-ph}\mathbf{P}_e. \quad (S7)$$

Note that the collision frequencies $\nu_{e-ph}$ and $\nu_{ph-e}$ are related but not equal, since the concentrations of phonons and electrons generally differ. If $N$ collisions per unit time occur in a unit volume, the exact balance relation $N = \nu_{e-ph} n_e = \nu_{ph-e} n_{ph}$ must hold, where $n_{ph}$ and $n_e$ are the phonon and electron concentrations, respectively. At room temperature, $n_{ph}$ exceeds $n_e$ by one to two orders of magnitude, consistent with the fact that the phonon lifetime is considerably longer than the electron mean free time.

To obtain an equation for the phonon energy density $U_l$, we multiply the kinetic equation (S2) by $\hbar\omega = \hbar q c_s$ and integrate over momenta:

$$\frac{\partial U_l}{\partial t} + c_s^2\, div\, \mathbf{P}_{ph} = S_e, \quad (S8)$$

where $S_e = \int s_e(q, \mathbf{r}, t)\hbar q c_s d^3\mathbf{q}$ is the integrated energy input from the electrons into the crystal lattice. The collision integrals $St_{th}\{f_0\}$ and $St_U\{\delta f\}$ do not contribute to Eq. (S8) because they conserve the total phonon energy. The term $c_s^2 div\, \mathbf{P}_{ph}$ describes the contribution of lattice thermal conductivity to heat transport. In most metals, the electronic thermal conductivity exceeds the lattice thermal conductivity by one to two orders of magnitude, so this term provides only a small correction to the heat transport by electrons, which is governed predominantly by $S_e$ and its evolution.

Equations (S7)–(S8) explicitly capture the emergence of a directed phonon flux that is delayed with respect to the heating of the phonon subsystem. As seen from Eq. (S7), the growth rate of the phonon momentum $\mathbf{P}_{ph}$ is proportional to the gradient of the lattice thermal energy $\mathbf{\nabla} U_l$, which itself rises and decays on picosecond timescales.

## B. Generation of internal fields and THz emission by a potential source

The problem of radiation from various gradient-type sources acting on conduction electrons near a metal surface has been solved in Refs. [1–5]. Below we summarize the main results of those works in a unified form.

Suppose that a potential force $\mathbf{F}(x, z, t)$ acts on the electrons per unit volume – for instance, the phonon drag force associated with inhomogeneous heating of the crystal lattice. The electrons in a metal can be described as a fluid (using hydrodynamic equations) because the electron–electron collision rate is very high, especially at high temperatures, where the mean free path drops to a few nanometers even in good conductors. The equation describing the low-frequency motion of electrons then takes the form

$$\frac{\partial \mathbf{v}}{\partial t} = -\frac{e\mathbf{E}}{m} + \frac{\mathbf{F}}{mn_e} - \frac{\nabla p_e}{mn_e} - \nu_e \mathbf{v}, \tag{S9}$$

where $\mathbf{v}$ is the directed electron velocity, $\mathbf{E}$ is the self-consistent electric field inside the metal, $\nu_e$ is the effective electron collision frequency, and $p_e$ is the electron pressure. In the ideal-gas model, $p_e = \frac{2}{3}n_e\varepsilon_e$, with $\varepsilon_e$ being the average kinetic energy per electron. At room temperature and above, $\nu_e$ is determined predominantly by collisions with phonons, so that $\nu_e \cong \nu_{e-ph}$. When electrons are heated to temperatures comparable with the Fermi energy, electron–electron Umklapp scattering also contributes noticeably to $\nu_e$.

The action of the force $\mathbf{F}$ induces a charge-separation field $\mathbf{E}$ and an electron density perturbation $\delta n_e$, related by Maxwell's equation

$$div\ \mathbf{E} = -4\pi e \delta n_e. \tag{S10}$$

Using the continuity equation

$$div\ n_e \mathbf{v} = -\frac{\partial \delta n_e}{\partial t} \tag{S11}$$

and calculating the divergence of Eq. (S9), we find:

$$\frac{\partial^2 \delta n_e}{\partial t^2} + \nu_e \frac{\partial \delta n_e}{\partial t} + \omega_p^2 \delta n_e = -\frac{div\ \mathbf{F}}{m} + \frac{\Delta p_e}{m}, \tag{S12}$$

where $\omega_p^2 = 4\pi n_e e/m$ is the plasma frequency and $\Delta$ is the Laplace operator. Here the background free-electron density $n_e$ is taken to be constant, which is justified for electron temperatures below 0.5–1 eV, that is, below the threshold of significant internal ionization.

Since the plasma frequency in metals $\omega_p \sim 10^{16}\ s^{-1}$ exceeds by orders of magnitude both the characteristic time of $\mathbf{F}$ variations and the collision frequency $\nu_e$, Eq. (S12) gives the following solution:

$$\delta n_e = -\frac{div\ \mathbf{F} + \Delta p_e}{m\omega_p^2} \cong \frac{1}{m\omega_p^2}\left(\frac{\partial^2 p_e}{\partial x^2} - \frac{\partial F_x}{\partial x}\right). \tag{S13}$$

Here we have also taken into account that the force $\mathbf{F}$ and the pressure $p_e$ are localized in a thin near-surface layer, so that $\frac{\partial}{\partial x} \gg \frac{\partial}{\partial z}$. Equation (S13) describes the electron density perturbation produced inside the metal by the force $\mathbf{F}$ and the electron pressure gradient, including their screening by charge-separation fields.

Because neither electron heat flux nor phonon flux crosses the metal surface, we have at the boundary $\frac{\partial p_e}{\partial x} = F_x = 0$. It follows that $\int_{-\infty}^{0} \delta n_e(x)dx = 0$, that is, the perturbation has the character of a distributed double layer with zero net surface charge. Accordingly, the normal component of the electric field of this layer is entirely localized within it, similar to the field of a parallel-plate capacitor.

Since the electron pressure and the phonon flux vary slowly compared with the plasma frequency, this double layer can emit radiation that leaves the surface. From Maxwell's equations we obtain

$$\nabla \, div \, \mathbf{E} - \Delta \mathbf{E} = -\frac{1}{c^2}\frac{\partial^2 \mathbf{E}}{\partial t^2} + \frac{4\pi}{c^2}n_e e \frac{\partial \mathbf{v}}{\partial t}. \qquad (S14)$$

Here we make use the expression for $\delta n_e$ since $div \, \mathbf{E} = -4\pi e \delta n_e(x,z,t)$. Taking into account Eq. (S9) and the relations $\omega_p^2 \gg \nu_e, \frac{\partial}{\partial t}$ and $\frac{\partial}{\partial x} \gg \frac{\partial}{\partial z}$, we find:

$$\frac{\partial^2 \mathbf{E}}{\partial x^2} - \frac{\omega_p^2}{\nu_e c^2}\frac{\partial \mathbf{E}}{\partial t} = -\frac{1}{n_e e}\left(\frac{\partial^2(\nabla p_e - \mathbf{F})}{\partial x^2} - \frac{\omega_p^2}{\nu_e c^2}\frac{\partial(\nabla p_e - \mathbf{F})}{\partial t}\right). \qquad (S15)$$

As is evident from Eq. (S15), the differential operators acting on $\mathbf{E}$ and on $(\nabla p_e - \mathbf{F})$ are identical. Hence, we obtain the quasistatic solution

$$\mathbf{E} = \frac{\mathbf{F} - \nabla p_e}{n_e e}. \qquad (S16)$$

The obtained solution corresponds to the quasistatic limit, i.e. zero electron velocity $\mathbf{v} = 0$ and $rot \, \mathbf{E} = 0$. This reflects the fact that, in the dense electron plasma of a metal, any potential force is almost immediately compensated by a charge-separation field.